\documentclass[compsoc,conference,a4paper,10pt,times]{IEEEtran}
\IEEEoverridecommandlockouts
\usepackage{cite}
\usepackage{amsmath,amssymb,amsfonts}
\usepackage{algorithmic}
\usepackage{graphicx}
\usepackage{textcomp}
\usepackage{bmpsize}
\usepackage{xcolor}
\usepackage{lipsum}
\usepackage{booktabs}
\usepackage[hyphens]{url}
\usepackage[colorlinks=true,urlcolor=black,breaklinks=true]{hyperref}
\usepackage{cleveref}
\usepackage{xspace}
\usepackage{caption}
\usepackage{subcaption}
\usepackage{balance}

\usepackage[binary-units]{siunitx}
\DeclareSIUnit{\nothing}{\relax}

\newcommand{\kn}{\kilo\nothing}
\newcommand{\sk}[1]{\SI{#1}{\kn}}

\usepackage{pifont} %
\newcommand{\cmark}{\textcolor{green}{\ding{51}}\xspace}%
\newcommand{\xmark}{\textcolor{red}{\ding{55}}\xspace}%

\def\BibTeX{{\rm B\kern-.05em{\sc i\kern-.025em b}\kern-.08em
    T\kern-.1667em\lower.7ex\hbox{E}\kern-.125emX}}

\newcommand{\ie}{i.e., \@}
\newcommand{\eg}{e.g., \@}

\newcommand{\cf}{cf. \@}

\newcommand{\etal}{et al.\xspace}
\newcommand{\one}{(1)~}
\newcommand{\two}{(2)~}
\newcommand{\three}{(3)~}

\newcommand{\sr}{segment routing\xspace}
\newcommand{\Sr}{Segment routing\xspace}

\usepackage[absolute,showboxes]{textpos}

\begin{document}

\setlength{\TPHorizModule}{\paperwidth}
\setlength{\TPVertModule}{\paperheight}
\TPMargin{5pt}
\begin{textblock}{0.8}(0.1,0.02)
    \noindent
    \footnotesize
    If you cite this paper, please use the WTMC reference:
    Victor-Alexandru Pădurean, Oliver Gasser, Randy Bush, Anja Feldmann. 2022.
    SRv6: Is There Anybody Out There?
    In \textit{International Workshop on Traffic Measurements for Cybersecurity 2022 (WTMC'22).}
\end{textblock}

\title{SRv6: Is There Anybody Out There?}

\author{\IEEEauthorblockN{Victor-Alexandru Pădurean}
\IEEEauthorblockA{\textit{MPI-INF} \\
vpadurea@mpi-inf.mpg.de}
\and
\IEEEauthorblockN{Oliver Gasser}
\IEEEauthorblockA{\textit{MPI-INF} \\
oliver.gasser@mpi-inf.mpg.de}
\and
\IEEEauthorblockN{Randy Bush}
\IEEEauthorblockA{\textit{Arrcus / IIJ} \\
randy@psg.com}
\and
\IEEEauthorblockN{Anja Feldmann}
\IEEEauthorblockA{\textit{MPI-INF} \\
anja@mpi-inf.mpg.de}
}

\maketitle

\begin{abstract}
\Sr is a modern form of source-based routing, \ie a routing technique where all or part of the routing decision is predetermined by the source or a hop on the path.
Since initial standardization efforts in 2013, \sr seems to have garnered substantial industry and operator support.
Especially segment routing over IPv6 (SRv6) is advertised as having several advantages for easy deployment and flexibility in operations in networks.
Many people, however, argue that the deployment of segment routing and SRv6 in particular poses a significant security threat if not done with the utmost care.

In this paper we conduct a first empirical analysis of SRv6 deployment in the Internet.
First, we analyze SRv6 behavior in an emulation environment and find that different SRv6 implementations have the potential to leak information to the outside.
Second, we search for signs of SRv6 deployment in publicly available route collector data, but could not find any traces.
Third, we run large-scale traceroute campaigns to investigate possible SRv6 deployments.
In this first empirical study on SRv6 we are unable to find traces of SRv6 deployment even for companies that claim to have it deployed in their networks.
This lack of leakage might be an indication of good security practices being followed by network operators when deploying SRv6.
\end{abstract}

\begin{IEEEkeywords}
segment routing, IPv6, SRv6
\end{IEEEkeywords}

\section{Introduction}

Since its initial specification, \sr (SR) has attracted attention from big vendors in the networking industry such as Cisco \cite{cisco-depl}, Juniper \cite{juniper-depl}, Huawei \cite{huawei-depl} \cite{srv6-deployments}, Nokia \cite{nokia-depl}, and Arrcus \cite{arrcus-depl}.
SR is a technique which leverages the source-routing principle, allowing a sender or an intermediate node to specify (in part) the path a packet takes on its way.
In addition, SR brings many promises \cite{cisco-sr}, one important benefit being the fact that there is no need to keep per-application and per-flow state, as all the necessary information is stored in the packet itself.
It also enables software-defined networking (SDN) and the selective use of different network appliances such as firewalls or snort~\cite{Abdelsalam2018DemoCO}.
Another benefit of SR mentioned by networking companies is its support of resilience techniques such as Topology Independent Loop-free Alternate Fast Re-route (TI-LFA) \cite{ietf-rtgwg-segment-routing-ti-lfa-08}, which in a failure scenario allows quickly shifting traffic to a backup path.

Moreover, many networking vendors claim to be working on adding SR-support into their products.
They also started offering support to their clients to integrate SR into their networks.
Such clients include SoftBank \cite{softbank-press}, Line Corporation \cite{line-corp}, MTN Uganda \cite{srv6-deployments}, Indosat Ooredoo \cite{indosat-press}, Rakuten \cite{rakuten-press}, Bell Canada \cite{bell-press}, China Unicom \cite{china-unicom-press}, China Telecom \cite{srv6-deployments}, Iliad \cite{iliad-press}, and CERNET2 \cite{srv6-deployments}.

Recently, \sr also started to attract the attention of the research community, with papers describing how new functions can be implemented on top of \sr \cite{tulumello2020micro} \cite{mayer2019network} \cite{paolucci2016interoperable} \cite{bhatia2015optimized}, review standardization activities, and review articles dedicated to \sr \cite{ventre2020segment}.

Furthermore, since SR is a relatively new technology compared to \eg MPLS \cite{rfc3031}, it is not yet as mature.
Thus, network operators engage in lively discussion about the benefits and possible downsides such as security implications of large-scale SR deployment, especially SR over IPv6 (SRv6) \cite{nanog-mails}.
Security issues may also arise from the uninformed use of SRv6 in a network \cite{ietf-topicbox}.
Moreover, the IETF community is also looking into security issues related to SRv6 deployment \cite{li-spring-srv6-security-consideration-07}.
Therefore, in this paper we aim to investigate the current state of SRv6 deployment in the wild.

\section{Background and Related Work}

\begin{figure}[t!]
\centerline{\includegraphics[width=0.9\columnwidth]{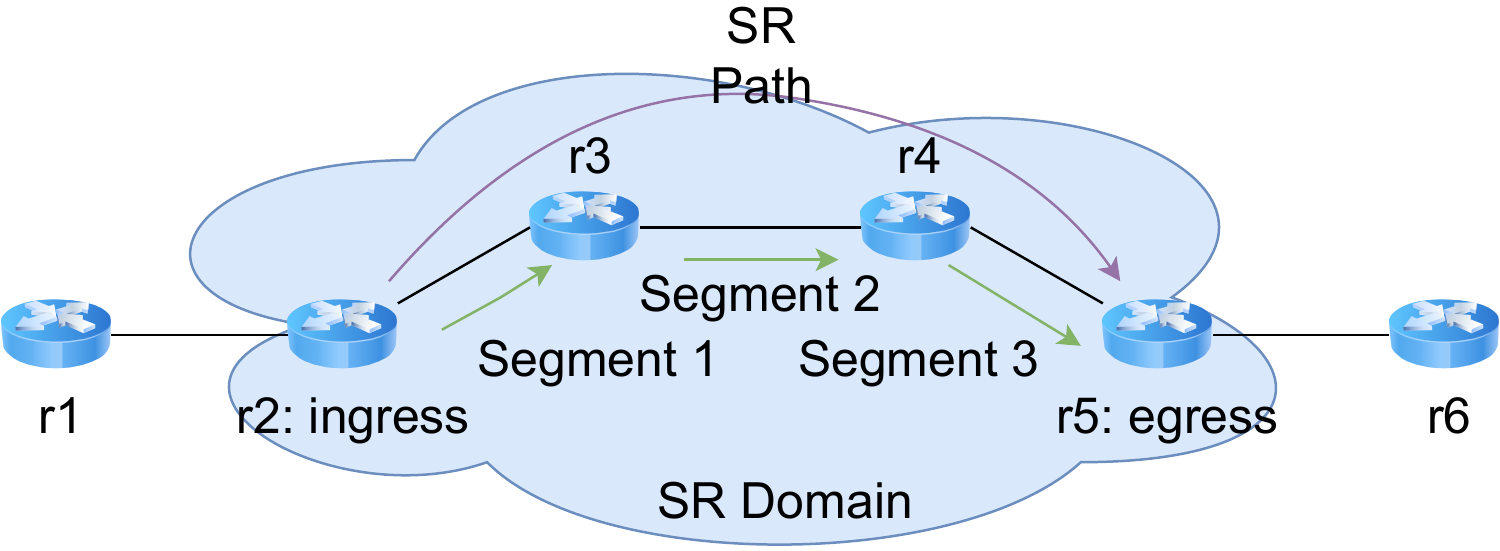}}
\caption{\Sr topology example showing SR domain, ingress, egress, SR path, and segments \cite{juniper-doc}.}
\label{fig:srv6topology}
\end{figure}

\Sr is based on the concept of source routing, \ie it allows a source node or an intermediate node to predetermine (part of) the path taken towards a destination \cite{rfc8402}.
\Sr, however, goes beyond a pure list of forwarding instructions:
It can chain services and obtain complex behaviors as a solution for service differentiation.
A \textit{segment} is the basis for \sr.
It is composed of a \textit{locator} (\ie a unique identifier of the network node where the instruction should be executed), an \textit{instruction} the node executes, either topological (\ie forwarding) or requiring a service to be executed, and optional \textit{arguments} for the instruction.
Segments can be recognized by their SR Segment Identifiers (SIDs).
We can chain segments into a list which connects the \textit{ingress node} (\ie headend, where the packet becomes ``SR-aware'') to the \textit{egress node} (\ie endpoint, either the final destination of the packet or the node where the SR capabilities are removed from a packet).
This list of segments is called an \textit{SR path}.
The set of nodes a packet travels between the ingress and egress nodes constitutes an \textit{SR domain}.
In \Cref{fig:srv6topology} we depict these concepts in an example topology.

Furthermore, an \textit{SR policy} describes how traffic is handled within an SR domain \cite{srpolicy}.
A policy is identified through the tuple \texttt{<headend, color, endpoint>}.
The color is used to differentiate between policies that have the same headend and endpoint.
Note that a policy can have multiple SR paths available (\eg for redundancy).
In this case a preference determines which SR policy is being picked.

In order to distribute SR policies we can make use of BGP update messages \cite{srbgp}.
BGP speakers can propagate candidate paths for SR policies by making use of Color Extended Communities.

\begin{figure}[t!]
\centerline{\includegraphics[width=0.9\columnwidth]{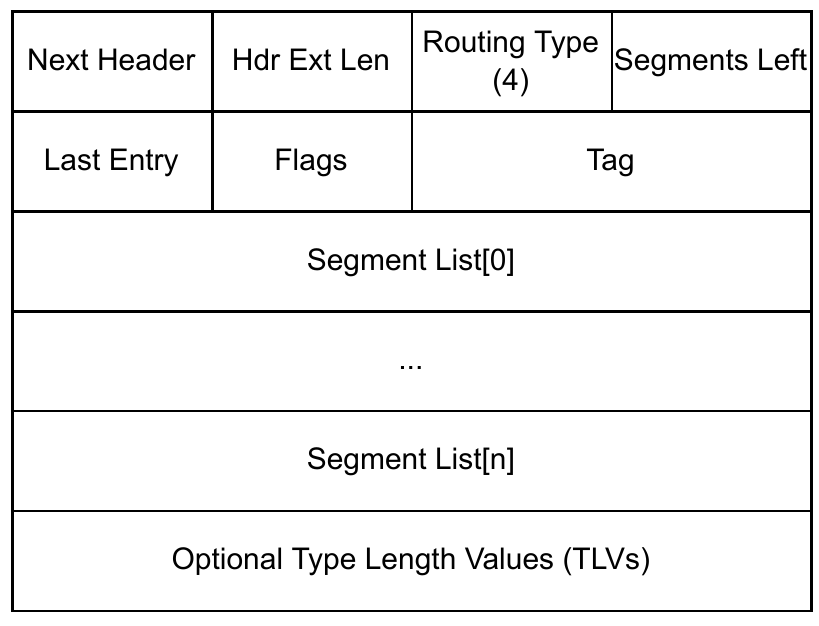}}
\caption{SRv6 header structure.}
\label{fig:srv6header}
\end{figure}

On the protocol level, \sr works on top of MPLS or IPv6.
Due to the long history of using MPLS in networking \cite{rfc3270} \cite{rfc2702} \cite{rfc3031} \cite{809383}, SR-MPLS is thought to be more mature than SR over IPv6 \cite{mpls-nanog}.
In this paper we focus on the prevalence of SR implemented on top of IPv6, \ie \textit{SRv6}.
This is done through a new type of routing header (next header value 43) called \textit{Segment Routing Header}, with routing type 4 \cite{rfc8754}.
The details regarding this header can be seen in \Cref{fig:srv6header}.
Further details of SRv6 are described in the corresponding RFC~\cite{rfc8986}.

Lately, researchers started exploring the capabilities and implications of \sr.

Ventre \etal conduct a comprehensive survey on the introduction, motivation, and evolution of SR, SR-MPLS, and SRv6 \cite{ventre2020segment}.
The authors provide a classification of current SR activities SR.
They conclude that SR's potential is not yet fully leveraged.
For example, they find a lack of works focusing on failure monitoring, but they seem confident that the network programming capabilities of SR will get more attention in the coming years.

Tulumello \etal propose ideas on how to make SRv6 more efficient by reducing the size of SIDs \cite{tulumello2020micro}.
Instead of representing only one segment with a single SID, they encode up to six micro instructions, represented through micro SIDs.
This drastically reduces the SIDs needed to be present in the SRH, thus reducing the packet overhead of SRv6 containing many SIDs.

Mayer \etal present ideas on how SRv6 can be integrated in an IoT and cloud infrastructure leading to a new distributed processing model \cite{mayer2019network}.
They theorize an abstract computing machine, which they call an SR-IoT Computing Machine which can be programmed with its own instruction set.
Thus, by building their own toolchain and leveraging the capabilities of \sr, the authors treat the whole Cloud-IoT network (i.e., without differentiation between core and edge) as a single logical machine.
This way a developer focuses on the application logic, while the owner of the infrastructure enforces management policies.

Lebrun and Bonaventure describe their experience implementing SRv6 in the Linux kernel \cite{lebrun2017implementing}.
In addition, they analyze the performance impact of SRH insertion.
They find that the performance overhead for SR entry points with their implementation for Linux 4.10 is limited to 15\%.

To the best of our knowledge this work is the first conducting an empirical analysis of the SRv6 landscape in the wild.

\section{Emulation}
\label{sec:emulation}

\begin{figure}[htbp]
\centerline{\includegraphics[width=0.9\columnwidth]{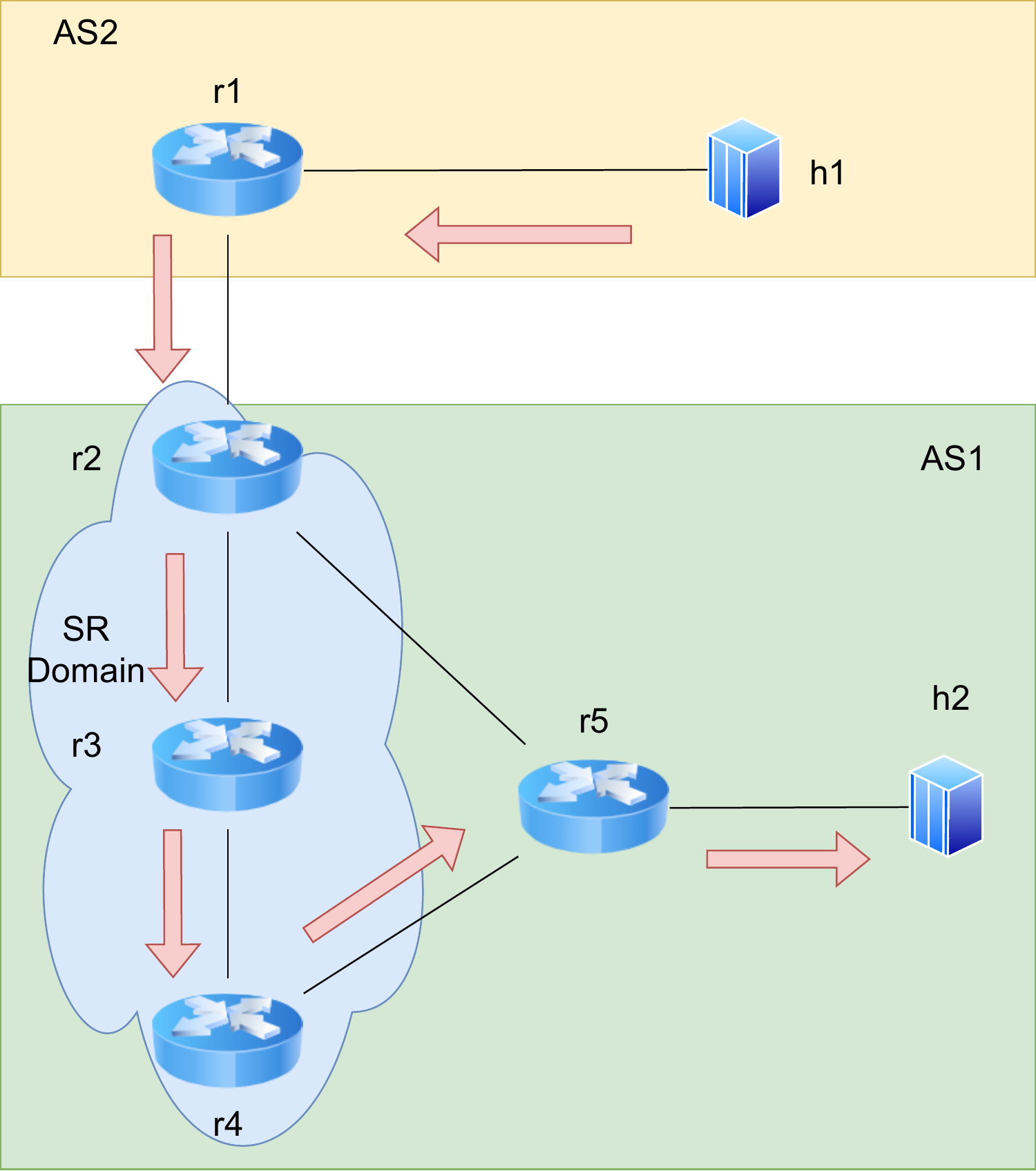}}
\caption{SRv6 emulation topology showing two ASes, one of which contains an SR domain with an SR path. The arrows show a packet's path.}
\label{fig:srv6emulation}
\end{figure}

Before analyzing the deployment of SRv6 in BGP and with traceroute measurements, we set up a controlled emulation environment in a lab.
We use IPMininet \cite{ipmininet}, a Python library that extends Mininet \cite{mininet}, enabling the use of SRv6 and BGP.
IPMininet directly uses the Linux kernel's implementation of \sr.
The setup is simple, yet complex enough to emphasize the traces that SRv6 may leave in an multi-AS environment.
We worked in an IPv6-only setting.
\Cref{fig:srv6emulation} shows our emulation topology.
We use one main AS---namely AS1---with 4 routers (\ie r2, r3, r4, and r5), forming an iBGP full mesh.
AS1 also contains one host system: h2, connected to r5.
Moreover, we set up a smaller AS---namely AS2---containing only one router (\ie r1).
There is also a host in AS2 (\ie h1) which is connected to the router r1.
The edge routers r1 and r2 are exchanging reachability information through an eBGP session.
Note that all routers run both OSPF and BGP.
Finally, we enable SRv6 on all the nodes and set up r2 as an SR entry point which enforces a simple SR policy: when receiving a packet destined to h2, r2 should forward it through r4.

This policy can be implemented by choosing one of the two SRv6 modes: \texttt{H.Insert} (called \texttt{inline} in the Linux Kernel implementation) or \texttt{H.Encap} (\texttt{encap} in Linux).
The \texttt{encap} mode indicates that the incoming packet will be encapsulated into another IPv6 packet (\ie IPv6-in-IPv6) at the ingress node (\ie r1).
The source of the outer packet is set to r2 and the destination address is set to the next segment (\ie r4 in our case, which is also our egress node).
Moreover, the entry point will add a segment routing header (SRH).
Note that in our example here, the segment r4 in the SRH can be omitted if a reduced SRH is being used \cite{rfc8754}.
The egress node will strip the packet of the outer IPv6 layer and forward it to the original destination (\ie h2).

The \texttt{inline} mode, however, aims to modify the IPv6 layer of the original packet.
It changes the destination address, adds an SRH, but leaves the source address untouched.
This means that an ICMPv6 message (\eg due to Hop Limit exceeded) will reach the original source \cite{lebrun2017implementing}, allowing detection of SRv6 deployment with traceroute.
Our findings confirm this.
We send \texttt{traceroute} probes from h1 to h2, capturing all the traffic at h1.
When using \texttt{encap}, no packet contained traces of SRv6, but when using \texttt{inline}, we can see SRHs in the ICMPv6 packets returned from the SR domain.
This is a promising finding, as various parties claim to implement and use \texttt{H.Insert} and \texttt{H.Insert.Red} (\ie reduced insertion) \cite{srdeployment}.

\textbf{To summarize:} Our emulation experiments show that it is possible to detect SRv6 deployment with traceroute, if SRv6 is implemented with inline mode.

\section{Investigating BGP Route Collectors}

One possible source of SRv6 leakage is BGP.
We inspect BGP data from 10 popular route collectors from RIPE RIS and RouteViews (see \Cref{tab:comms_rcs} for a list of RCs) in September 2021.

We find that Color Extended BGP Communities \cite{rfc5512} can be used to steer traffic according to various SR policies \cite{colorcisco,srbgp}.
Thus we explore data from BGP archives looking for these types of communities.
This can be done by detecting the specific bytes in the Type and Sub-Type fields of extended communities: \texttt{0x03} and \texttt{0x0b}.
Our search for these byte combinations does not yield results in the BGP collector data.

Next, we also search for other communities that may be an indicator of SRv6 usage, directly in the BGP data.
Then, we collect a list of ASes that belong to organizations that claim to be using SR (we will call them SR-suspect ASes).
We then follow a simple procedure for inspecting whether these SR-suspect ASes have specific communities in common.
First, we extract all BGP communities that appear with at least one SR-suspect AS on the AS path.
Subsequently, we extract the communities that never appear with SR-suspect ASes on the path.
Finally, we make a difference of both of these sets of BGP communities and obtain a list of communities which only appear in BGP announcements with SR-suspect ASes on the path.
We then count in how many SR-suspect ASes each community appears, aiming to find a community used by as many SR-suspect ASes as possible.
However, each community we find is only used by a small number of ASes.
As can be seen in \Cref{tab:comms_rcs}, the majority of communities is used by only a single AS, whereas a small number is used by two or more ASes.
We further inspect the cases where BGP communities are seen in announcements with more than one SR-suspect AS.
We find that these cases are almost all related to sibling ASes, \ie two ASes with a different ASN but operated by the same organization.
Only at \texttt{rv4} we find three communities within a single BGP announcement with two suspect ASes on the AS path.
We find that those communities are related to tagging \cite{softbankcomm} and peer selection \cite{nttcomm}, and not SRv6.

\begin{figure}[t!]
\centerline{\includegraphics[width=0.8\columnwidth]{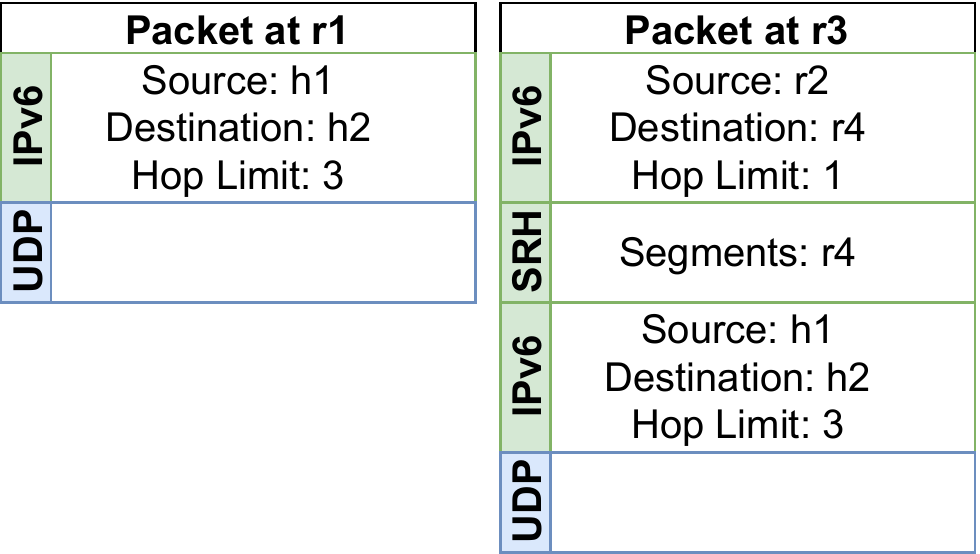}}
\caption{Packet structure and content for \texttt{H.Encap} behavior at routers r1 and r3.}
\label{fig:encap}
\end{figure}

\begin{figure}[t!]
\centerline{\includegraphics[width=0.8\columnwidth]{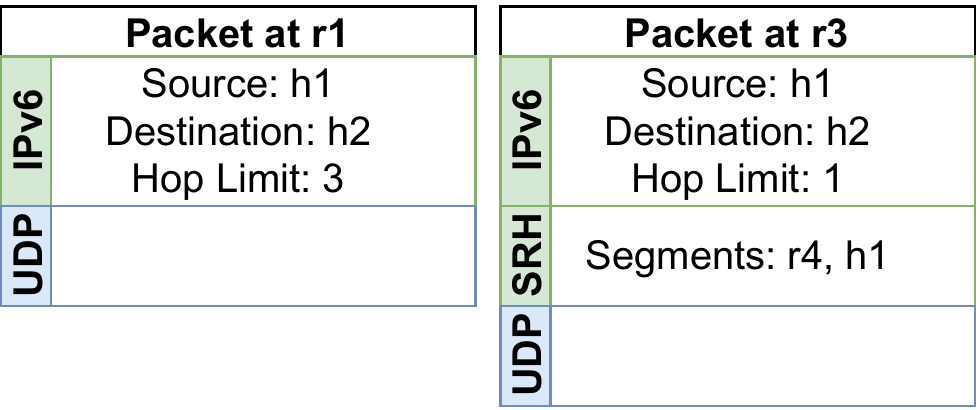}}
\caption{Packet structure and content for \texttt{H.Insert} behavior at routers r1 and r3.}
\label{fig:inline}
\end{figure}

Moreover, we investigate whether the communities found only for SR-suspect ASes are location communities \cite{silva2022automatic}.
We find that between 25\% to 50\% of all communities from \Cref{tab:comms_rcs} are identified by the BGP location community database \cite{locationcomm} as location communities.
This shows that a non-negligible portion of these communities is very unlikely to be used for SR.

Furthermore, we inspect BGP path attributes as potential indicators of SR deployment.
These include \texttt{BGP Prefix-SID}, \texttt{BGP-LS Attribute} and \texttt{Tunnel Encapsulation}.
The \texttt{BGP Prefix-SID} attribute \cite{rfc8669} is a BGP extension used for signaling information about BGP Prefix Segment Identifiers, thus it is a good indicator of SR deployment.
\texttt{BGP-LS} \cite{rfc7752}, on the other hand, is used for more general purposes, yet it has been extended \cite{rfc9085} to support carrying SR information between ASes via BGP.
This also applies to the \texttt{Tunnel Encapsulation} attribute \cite{rfc9012}, which permits carrying Prefix-SID-related information in one of its Sub-TLVs.
We search the BGP collector archives for interesting path attributes, but we only find generic attributes (\eg \texttt{AGGREGATOR}, \texttt{ATOMIC\_AGGREGATE}, \texttt{AS\_PATHLIMIT}) and no SR-specific attributes.

\begin{table}[!t]
    \centering
    \caption{Unique BGP community values seen only at suspect ASes (\# comms), seen only at multiple suspect ASes (\# multiple comms), and seen only at multiple suspect ASes while ignore sibling communities (\# non-sibling comms).}
    \label{tab:comms_rcs}
    \resizebox{\columnwidth}{!}{
    \begin{tabular}{lrrr}
       \toprule
       Collector & \# comms & \# multiple comms & \# non-sibling comms \\
       \midrule
       rv2       & 208      & 14                & 0 \\
       rv4       & 257      & 23                & 3 \\
       rv5       & 0        & 0                 & 0 \\
       rv6       & 97       & 9                 & 0 \\
       rv-amsix  & 190      & 8                 & 0 \\
       rrc00     & 257      & 16                & 0 \\
       rrc01     & 206      & 11                & 0 \\
       rrc04     & 151      & 5                 & 0 \\
       rrc05     & 171      & 8                 & 0 \\
       rrc06     & 89       & 1                 & 0 \\
       \bottomrule
    \end{tabular}
    }
\end{table}

\textbf{To summarize:} We analyze data collected by 10 BGP collectors, looking for attributes that could indicate SRv6 deployment.
We find no such attributes.
Furthermore, we conduct a thorough investigation of BGP Communities, especially Color Extended Communities, aiming to find a pattern that may indicate SRv6 usage.
Again, we do not find any relevant signal.
This means that these attributes are either properly filtered by BGP speakers on the path \cite{krenc2021level} or the egress routers of SRv6 deployments.

\section{Tracerouting for SRv6}

In order to reveal SRv6 deployment in the Internet, we ran several traceroute measurements.
Further, we want to understand if we can identify SRv6 in networks which claim to have deployed it \cite{srdeployment}.
Companies claiming to deploy \texttt{H.Insert} and its variations are especially valuable to us, since those might leak information in traceroute (\cf \Cref{sec:emulation}).
Those companies are Iliad, SoftBank, Line Corporation, China Unicom, China Telecom, and CERNET2.
We use PeeringDB \cite{peeringdb} and BGPView \cite{bgpview} to collect AS numbers related to these organizations.
We then use the WHOIS database and CAIDA's Routeviews BGP data \cite{routeviews} to get IPv6 prefixes for the identified ASes.
Finally, we generate 100 random addresses for each prefix resulting in \sk{213.8} addresses in total.

Next, we use Yarrp \cite{beverly2016yarrp} to run traceroute towards each of these addresses.
Before conducting active measurements we incorporate proposals by Partridge and Allman \cite{partridge2016ethical} and Kenneally and Dittrich \cite{kenneally2012menlo} and follow best measurement practices \cite{durumeric2013zmap} by using dedicated servers, informative rDNS names, a website with information, and maintaining a blocklist.
We did not receive any complaints while conducting the measurements.
We run the measurements from a single vantage point at MPI-INF, which might influence the traceroute paths we cover.
We send two types of probes: SR-unaware probes and SR-enabled probes.
The former are simply regular traceroute probes, the latter sends probes containing a \sr header.
We modify Yarrp in order to send SR-enabled probes.
Sending SR-enabled probes allows us to check if SRv6-enabled routers modify the list of segments in the SRH, which would be visible in the returned quoted ICMPv6 packet.
We conduct both types of measurements for various transport protocols, as shown in \Cref{tab:traceroutes}.
We target three types of addresses:
\one We take a random sample of 10 million addresses from the IPv6 Hitlist \cite{gasser2018clusters} (marked as \textit{hitlist} in \Cref{tab:traceroutes}). \two We generate random addresses within prefixes of SR-suspect ASes (marked as \textit{suspect} in {tab:traceroutes}).
\three We generate random addresses for each BGP-announced prefix.
Even though we see SRHs being returned from the SR-enabled measurements, those are not modified at all and therefore do not leak any SRv6 deployment.
Moreover, our other measurements do not return any SRH and therefore do not leak SRv6 deployment in the wild.

\begin{table}[!t]
    \centering
    \caption{Overview of traceroute measurements.}
    \label{tab:traceroutes}
    \begin{tabular}{llclc}
       \toprule
       Date & Type & SRH sent & Targets & SRv6 leaked \\
       \midrule
       2021-11-09 & TCP SYN & \xmark & hitlist & \xmark \\
       2022-02-08 & TCP6 SYN & \xmark & suspect & \xmark \\
       2022-02-09 & TCP6 SYN & \cmark & suspect & \xmark \\
       2022-02-10 & ICMPv6 & \xmark & prefix & \xmark \\
       2022-02-10 & TCP6 ACK & \xmark & prefix & \xmark \\
       2022-02-10 & TCP6 SYN & \xmark & prefix & \xmark \\
       2022-02-10 & UDP6 & \xmark & prefix & \xmark \\
       2022-02-15 & ICMPv6 & \xmark & prefix & \xmark \\
       2022-02-15 & TCP6 ACK & \xmark & prefix & \xmark \\
       2022-02-17 & TCP6 ACK & \xmark & prefix & \xmark \\
       \bottomrule
    \end{tabular}
\end{table}

\textbf{To summarize:} We conduct traceroute measurements to identify SRv6 deployments in the wild, but do not see any traces of SRv6.
Either companies do not use SRv6's inline mode as they claim or they filter SRHs from the returned ICMPv6 packets.

\section{Conclusion}

\Sr is a protocol which has its roots in the source routing space.
It brings many promises, including SDN-readiness, fast rerouting, and statelessness.
Given its relatively young age, experts are concerned to the security implications it brings.
Therefore, in this paper we tried to measure the SRv6 deployment leakage of \sr in the wild.
We searched for SRv6 in the real world, by exploring BGP collector archives and actively probing addresses of organizations that claim to be using SRv6.
While analyzing the BGP data, we looked for path attributes that may be indicators of \sr and tried to find a correlation between communities and \sr usage.
Unfortunately, we found no trace of \sr.
BGP speakers may be filtering such attributes, or egress routers themselves may carefully pick what kind and which part of announcements are propagated to neighbors.
Finally, we conducted a traceroute measurement campaign in order to identify leaked SRv6 deployments in the Internet, but we could not find any trace.
Organizations either do not use the inline mode as they claim or they carefully filter \sr traces from the returned packets.
Another possibility is that operators might use SRv6 exclusively for 5G deployments, which are likely to be more firewalled than other types of networks, possibly resulting in fewer SRv6 leaks.
We release data \cite{3.19HAUU_2022} and analysis code \cite{codegithub} to be used by fellow researchers looking into investigating SRv6 deployment in the future.

\noindent\textbf{Future work:}
We will continue to leverage BGP collector data and traceroutes in the future to see if SRv6 leaks occur.
Moreover, we plan to extend the BGP analysis to cover more route collectors and longer timespans.
Additionally, we will increase the target set of our traceroute analysis to discover well hidden SRv6 deployments.
Finally, we plan to set up a hardware testing lab in order to extend our emulation to bare-metal devices.

\noindent\textbf{Acknowledgments:}
The authors acknowledge the partial financial support by the Federal Ministry of Education and Research of Germany in the program of ``Souverän. Digital. Vernetzt.'' joint project 6G-RIC (16KISK027).

\balance

\bibliographystyle{IEEEtran}
\bibliography{IEEEabrv,paper}

\end{document}